\documentclass[a4paper,scale=0.8,onecolumn,nobibnotes,nofootinbib]{revtex4}

\usepackage{amsmath,amssymb}
\usepackage{graphicx,subfigure,epsfig}
\usepackage{hyperref}
\usepackage{geometry}
\hypersetup{
    colorlinks=true,
    citecolor=blue,
    linkcolor=red,
    filecolor=magenta,
    urlcolor=cyan,}

\newcommand{\be}{\begin{equation}}
\newcommand{\ee}{\end{equation}}

\begin{document}

\title{Inclusive diffractive heavy quarkonium photoproduction \\in $pp$, $pA$ and $AA$ collisions}
\author{Yi Yang$^{1,}$\footnote{\textsl{E-mail address:} gs.yangyi17@gzu.edu.cn
}, Shaohong Cai$^{1,}$\footnote{\textsl{E-mail address:} caish@mail.gufe.edu.cn}, Yanbing Cai$^{2,}$\footnote{\textsl{E-mail address:} yanbingcaigufe@gmail.com}, Wenchang Xiang$^{1,2,}$\footnote{\textsl{E-mail address:} wxiangphy@gmail.com}}
\affiliation{$^1$Department of Physics, Guizhou University, Guiyang, 550025, China\\
$^2$Guizhou Key Laboratory in Physics and Related Areas, Guizhou University of Finance and Economics, Guiyang, 550025, China
}


\begin{abstract}
The inclusive $J/\Psi$ production by direct and resolved photoproduction in the $\gamma p$ scattering is calculated based on the nonrelativistic quantum chromodynamics (NRQCD) factorization formalism, which is in good agreement with the experimental data of total cross section distribution of heavy quarkonium production at HERA. Then we extend the formalism including the direct and resolved photoproduction processes to resolved pomeron model to study the heavy quarkonium photoproduction at the LHC energies. We present the predictions of rapidity and transverse momentum distributions of the inclusive diffractive $J/\Psi$, $\Psi(2S)$ and $\Upsilon$ photoproduction in $pp$, $pPb$ and $PbPb$ collisions at the LHC energies. Our numerical results indicate that the resolved photoproduction processes play an important role in the heavy quarkonium production. Especially for $pp$ collisions, the contribution of resolved photoproduction processes is the largest, which can reach to $28\%$, $13\%$ and $44\%$ for the rapidity distributions of $J/\Psi$, $\Psi(2S)$ and $\Upsilon$ inclusive diffractive photoproduction, respectively.

\end{abstract}
\maketitle
~~~~~~~~~\textsl{Keywords:} NRQCD factorization; Heavy quarkonium; Resolved photoproduction

\section{Introduction}
\label{sec:intro}
In recent few years, the study of heavy quarkonium has become an active field in high-energy physics, because it is an ideal tool to investigate the perturbative and non-perturbative regimes of quantum chromodynamics (QCD). Many experimental groups have released their experimental data about the inclusive $J/\Psi$ photoproduction, such as the ALICE, LHCb and CMS \cite{ALICE2011, LHCb2011, CMS2011}. In particular, the $pp$ collider with higher center-of-mass energy is constructing in the LHC. Therefore a great number of data of heavy quarkonium production can be used to check the theoretical predictions and motivate scientists to explore the mechanism of heavy quarkonium production.

There are kinds of models to describe the heavy quarkonium production, such as the vector meson dominance (VMD) model \cite{Rebyakova:2011vf, Guzey:2013xba, Frankfurt:2015cwa, Guzey:2016qwo}, the color-evaporation model (CEM) \cite{CEM1, CEM2, CEM3, CEM4, CEM5}, the color-singlet model (CSM) \cite{CSM1,CSM2,CSM3,CSM4} and the nonrelativistic QCD (NRQCD) factorization formalism \cite{NRQCD1, NRQCD2, doctorcai132}. The VMD model assumes that the photon fluctuates into vector mesons which subsequently can interact with hadrons \cite{Bauer:1977iq}. The VMD model to explain the behavior of the pion form factor and some features of the nucleon form factors at small momentum transfers is successful. The CEM assumes that the quarkonium can be produced when the invariant mass of the quark pair is less than the threshold of open-flavor heavy mesons. The CEM mechanism can explain the strong production processes of $J/\Psi$, but it cannot describe the transverse momentum distributions of heavy quarkonium production \cite{prd94 114029(2016)}. For the CSM factorization formalism, it assumes the heavy quark-antiquark pair ($c\bar{c}$ or $b\bar{b}$) has the same spin and angular-momentum quantum numbers, and the quark-antiquark pair which evolves into the heavy quarkonium is in a color-singlet state. The CSM mechanism is successful for quarkonium production at low energies \cite{Schuler:1994hy}. The inclusive $J/\Psi$ and $\Upsilon$ photoproduction have been computed by Goncalves and Moreira using the CSM mechanism, and their results can describe the inelastic $J/\Psi$ photoproduction in inclusive $\gamma p$ interactions at HERA \cite{Goncalves:2013}. Unfortunately, the CSM is inconsistent with the production of higher-orbital-angular-momentum quarkonium states \cite{CSM:chanllenge}. In the NRQCD factorization formalism, the cross section of the quarkonium includes the short-distance part and long-distance part. The short-distance part describes the production of quark pair ($c\bar{c}$ or $b\bar{b}$) and gluons systems with certain quantum numbers, which is calculated by the perturbative QCD. The long-distance part describes the systems composed of quark pair and gluons evolve into the bound states. Subsequently, the bound states will hadronize into the final state quarkonium. The long-distance part is expressed by long-distance matrix elements (LDME) $\langle \mathcal{O}^{V}[n]\rangle$ obtained via fitting the experimental data. The NRQCD factorization formalism has been successfully applied to study the production of the heavy quarkonium. Then we will study the photoproduction of heavy quarkonium based on the NRQCD factorization formalism to deepen our understanding about the mechanism of heavy quarkonium production.

It is well known that the photoproduction processes are important for the electron-proton ($ep$) deep inelastic scattering (DIS) \cite{epdis}. In the $ep$ DIS, a high-energy photon radiated from the electron can directly interact with the proton, which is named the direct photoproduction processes. The $\eta_c$ direct photoproduction, by the double-photon, inclusive and diffractive photon-hadron interactions, in $pp$ and $pPb$ collisions at LHC energies has been discussed using the NRQCD factorization formalism \cite{Goncalves:2018yxc}, and S. R. Klein in Ref. \cite{Klein:2018ypk} points out that the cross sections of the $J/\Psi \rightarrow \gamma \eta_c$ are larger than the cross sections of exclusive $\eta_c$ studied by Goncalves and Moreira. As we know, the inclusive events refer to the fact that the particle is produced by hadron breakup on one side. In the exclusive events, nothing else is produced apart from the primary hadrons and central quarkonium. The diffractive events are characterized by the intact hadrons in the final state. Our motivation in this work is to study the inclusive diffractive $J/\Psi$, $\Psi(2S)$ and $\Upsilon$ production in $pp$, $pPb$ and $PbPb$ collisions.

Moreover, due to the uncertainty principle, the high-energy photon in the $ep$ DIS will fluctuate out a quark-antiquark pair to interact with the partons of the proton, which is named the resolved photoproduction processes. Especially, the photon in the resolved photoproduction processes is regarded as a kind of hadron-liked particle that consist of the quarks and gluons. Ingelman and Schlein proposed that the pomeron also has a partonic structure \cite{pomeron:partonic_structure}, and the resolved pomeron model can be used to describe the quarkonium production. In the Ref. \cite{prc Fu.Y.P} the resolved photoproduction processes have considerable contribution for production of large transverse momentum dileptons at LHC energies. Therefore, to explore the contribution of the resolved photoproduction processes for heavy quarkonium production, we will consider to combine the resolved photoproduction processes with the resolved pomeron model to investigate the $J/\Psi$, $\Psi(2S)$ and $\Upsilon$ inclusive diffractive photoproduction in $pp$, $pPb$ and $PbPb$ collisions.

The paper is organized as follows. In the next section we will introduce the formalism of heavy quarkonium production with the photoproduction processes and resolved pomeron model. In Sec. \ref{sec:result}, the numerical results are presented. Eventually, the summary is given in Sec. \ref{sec:summary}.

\begin{figure}[t!]
\begin{center}
\hspace{-0.3cm}
\includegraphics[scale=0.5]{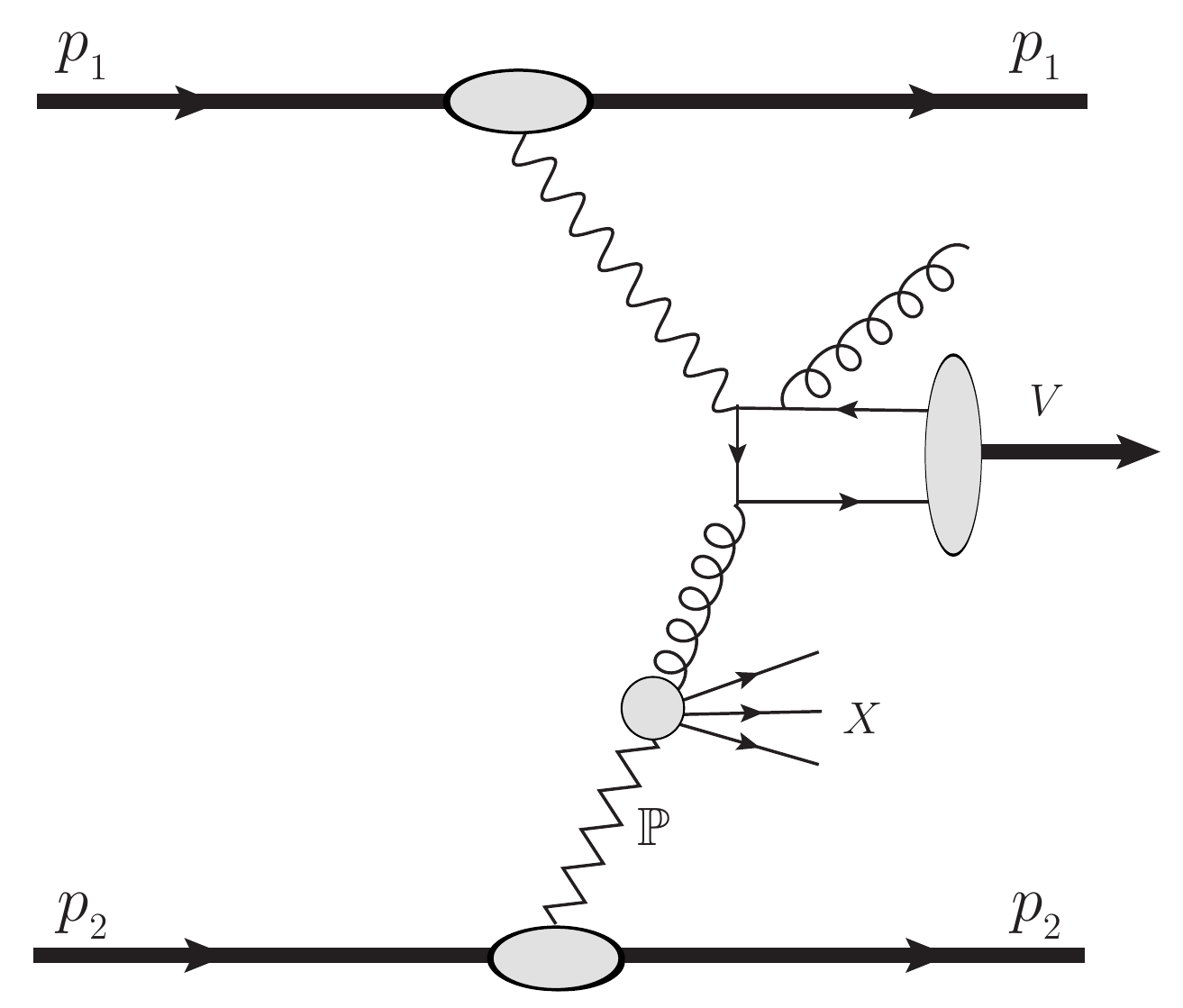}
\hspace{0.5cm}
\includegraphics[scale=0.5]{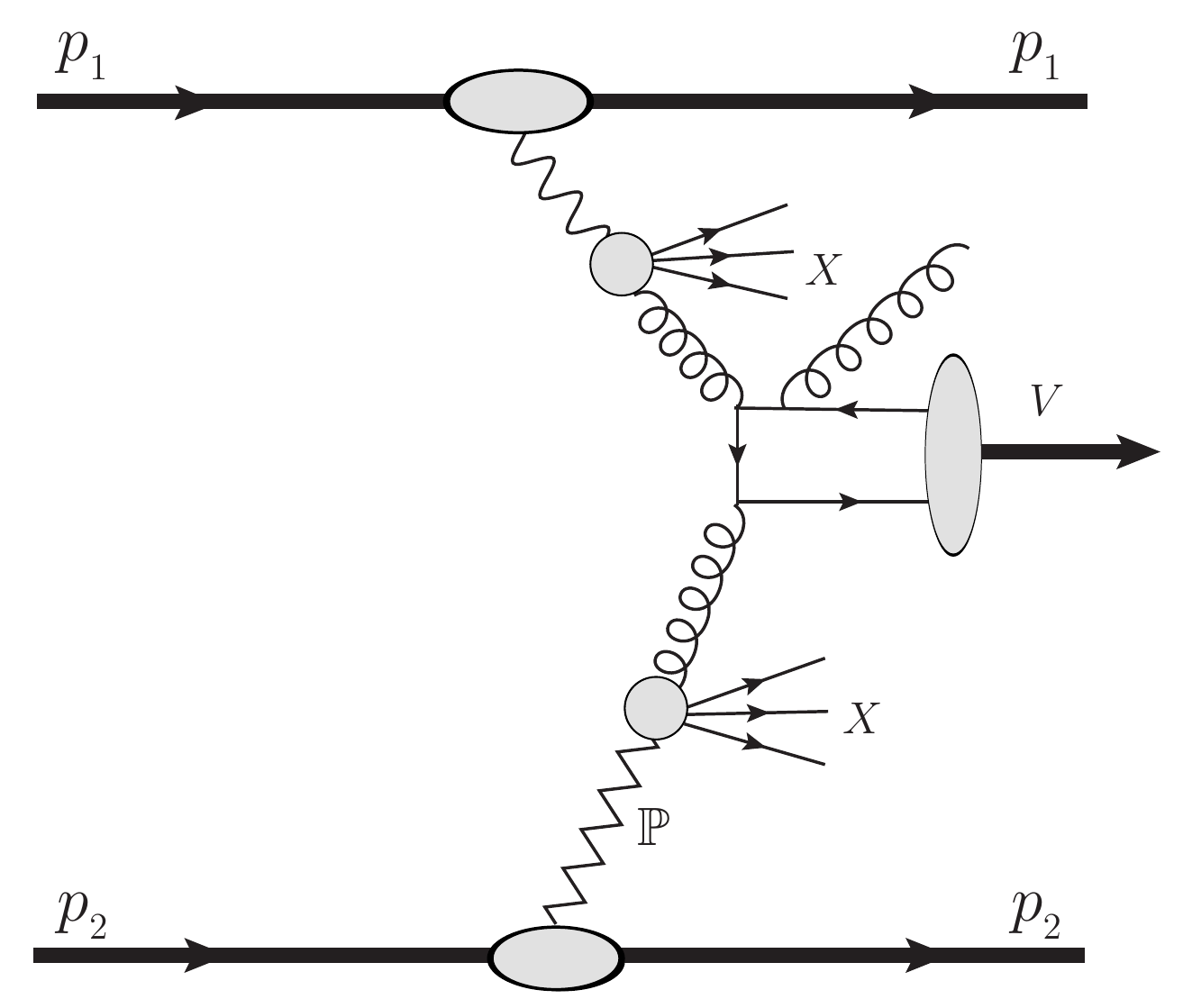}
\end{center}
\setlength{\abovecaptionskip}{0.1cm}
\setlength{\belowcaptionskip}{1cm}
\caption{The schematic diagrams of inclusive diffractive heavy quarkonium direct (left panel) and resolved (right panel) photoproduction with the resolved pomeron model.}
\label{feynman}
\end{figure}

\section{The cross section of heavy quarkonium photoproduction in NRQCD}\label{sec:two}

\subsection{Total cross section and rapidity distribution of heavy quarkonium photoproduction}
The schematic diagrams of inclusive diffractive heavy quarkonium photoproduction are presented in Fig. \ref{feynman}. As illustrated in Fig. \ref{feynman}, we consider the direct photoproduction processes (left panel in Fig. \ref{feynman}) and resolved photoproduction processes (right panel in Fig. \ref{feynman}) to study heavy quarkonium production. In the direct photoproduction processes ($\gamma g$  processes) the pomeron emitted from the hadron ($p_2$) can radiate a gluon to interact with photon emitted from another hadron ($p_1$). In the resolved photoproduction processes ($gg$ processes), a high-energy hadron ($p_1$) will radiate a photon, and owing to the uncertainty principle the photon can fluctuate out a gluon to interact with the gluon radiated from the resolved pomeron. In these processes, the total cross section will be factorized into two parts: one, the equivalent photon flux of the relativistic hadron, and the second, the photon-hadron cross section. For the heavy quarkonium photoproduction, its cross section is given by \cite{cpl32,cpl}
\be\label{eq1}
\sigma (p_1+p_2 \rightarrow p_1 \otimes V+X \otimes p_2)= \int d \omega \frac{dN_{\gamma /p_1}(\omega)}{d\omega}\sigma _{\gamma p_2 \rightarrow VX\otimes p_2}(W_{\gamma p_2})+\int d\omega\frac{dN_{\gamma /p_2}(\omega)}{d\omega} \sigma_{\gamma p_1 \rightarrow VX\otimes p_1}(W_{\gamma p_1}),
\ee
where $\otimes $ represents the presence of a rapidity gap in the final state, $\omega$ is the photon energy in the proton-proton collision system and $W_{\gamma p}$ is the photon-hadron center-of-mass energy. $W_{\gamma p}$ and $\omega$ are given by
\be\label{eq2}
W_{\gamma p}=\sqrt{2\omega \sqrt{s}}, \quad \omega =\frac{M}{2}exp(\pm y),
\ee
where$\sqrt{s}$ is the center-of-mass energy of the proton-proton collision system and $M$ is quarkonium's mass. In Eq. (\ref{eq1}), $dN_{\gamma /p_i}/d\omega$ is the equivalent photon flux of the relativistic proton $p_i$ ($i$=1, 2), which is given by \cite{cpl33, Goncalves:2012cy,photon spectrum, Goncalves:2017zdx}
\be
\frac{d N_{\gamma /p}(\omega )}{d\omega }=  \frac {\alpha _{em}}{2\pi \omega }\Big[1 + (1 -\frac{2\omega}{\sqrt{s}})^2\Big]\times \Big( \ln{\eta} - \frac{11}{6} + \frac{3}{\eta}  - \frac{3}{2\eta ^2} + \frac{1}{3 \eta ^3}\Big),
\ee
where $\eta=1+\Big[(0.71\,GeV^2)/Q_{min}^2\Big]$, and $Q_{min}^2$ is equal to $(\omega/\gamma_L)^2$ at high energy limit.

Using the relation between $y$ and $\omega$ in Eq. (\ref{eq2}), the rapidity distribution of heavy quarkonium inclusive diffractive photoproduction is given by
\be\label{eq4}
\frac{d\sigma (p_1+p_2 \rightarrow p_1 \otimes V+X \otimes p_2)}{dy}=\omega \frac{dN_{\gamma / p_1 }(\omega )}{d\omega} \sigma _{\gamma p_2 \rightarrow VX\otimes p_2}(\omega )+ \omega \frac{dN_{\gamma /p_2}(\omega )}{d\omega} \sigma_{\gamma p_1 \rightarrow VX\otimes p_1}(\omega ),
\ee
where the two terms in Eq. (\ref{eq1}) and Eq. (\ref{eq4}) denote the photon emitted either from the projectile proton ($p_1$) or from the target proton ($p_2$). Therefore, the rapidity distribution of $pp$ collisions will be symmetric about the central rapidities ($y=0$). Here $\sigma _{\gamma p_2 \rightarrow VX\otimes p_2}$ and $\sigma_{\gamma p_1 \rightarrow VX\otimes p_1}$ are the total cross sections of photon-proton processes, which can be expressed by the NRQCD factorization formalism, and be discussed in the next two subsections.

\subsection{The cross section of direct photoproduction processes in NRQCD }
The calculation of the total cross section ($\sigma _{\gamma p \rightarrow VX\otimes p}$) is a key ingredient for investigating quarkonium photoproduction. In terms of the NRQCD factorization formalism, the total cross section can be factorized into the parton distribution function and partonic differential cross section. Thus the total cross section of heavy quarkonium photoproduction can be written as \cite{gammah,Goncalves:2017bmo}
\be\label{eq5}
\sigma (\gamma+p \rightarrow V+X\otimes p)=\int dzdp_T^2\frac{x_1 g_h(x_1,Q^2)}{z(1-z)}\frac{d\sigma}{dt}(\gamma+g\rightarrow V+g).
\ee

The variable $z = (p_q.p_h)/(p_\gamma.p_h)$ is the fraction of the photon energy carried by the quarkonium, where $p_q$, $p_h$ and $p_{\gamma}$ is the four momentum of the quarkonium, hadron and photon, respectively. Moreover $x_1$ is the momentum fraction of hadron carried by the gluon in the direct photoproduction processes, which can be written as
\be
x_1=\frac{p_T^2+M^2(1-z)}{W_{\gamma p}^2z(1-z)}.
\ee

In addition, $p_T$ is the quarkonium transverse three-momentum and the $d \sigma /dt$, which can be expressed by the long-distance matrix elements and Mandelstam variables $s$, $t$ and $u$, is partonic differential cross section of the heavy quarkonium photoproduction. In our calculations the differential cross section includes the contributions of color-singlet and color-octet states \cite{gammah, octet:dsdt}. The Mandelstam variables for the direct photoproduction processes ($\gamma +p\rightarrow V+X\otimes p$) are expressed as \cite{Goncalves:2013}
\begin{eqnarray}
s &=& \frac{p_T^2+M^2(1-z)}{z(1-z)}, \notag \\
t &=& -\frac{p_T^2+M^2(1-z)}{z}, \notag \\
u &=& -\frac{p_T^2}{(1-z)}.
\end{eqnarray}

In Eq. (\ref{eq5}), the $g_h$ is the diffractive gluon distribution, which can be expressed by the convolution of the pomeron distribution of the proton and the gluon distribution of the pomeron \cite{pomeron model}
\be
g_{h}(x,Q^2)=\int_x^1 \frac{dx_{\mathbb{P}}}{x_{\mathbb{P}}} f_{\mathbb{P}/p}(x_{\mathbb{P}})
 g_{\mathbb{P}}\Big(\frac{x}{x_{\mathbb{P}}},Q^2\Big),
\ee
with
\be
f_{\mathbb{P}/p}(x_{\mathbb{P}})=\int_{-1}^{t_{max}} \frac{\lambda e^{\beta t}}{x_{\mathbb{P}}^
{2\alpha_{\mathbb{P}}(t)-1}}dt,
\ee
where $\alpha_\mathbb{P}(t)=\alpha_\mathbb{P}(0)+\varepsilon t$, the slope of the pomeron flux $\beta=5.5 \, GeV^{-2}$ and $\varepsilon=0.06 \, GeV^{-2}$ is obtained by fitting the H1 data \cite{H1data}. The Regge trajectory of the pomeron is $\alpha_\mathbb{P}(0)=1.111\pm0.007$. The $t_{max}$ is equal to $-m_p^2x_\mathbb{P}^2/(1-x_\mathbb{P})$, and the proton mass is  $m_p=0.938\,GeV$. The normalization factor is $\lambda=1.7101$ \cite{Goncalves:2017bmo}. In our study, we use the parametrization of gluon distribution in pomeron ($g_{\mathbb{P}}$) in Ref. \cite{H1data}.

\subsection{The cross section of resolved photoproduction processes in NRQCD}

We combine, for the first time, the resolved pomeron model with the resolved photoproduction processes to investigate the heavy quarkonium production. In the resolved photoproduction processes, the photon is regarded as a kind of particle that is similar to hadron. Therefore, the photon will radiate the parton to interact with the parton emitted from the target hadron. It's known that the gluon distribution is much larger than the quark distribution in photon \cite{photon structure} and hadron \cite{cteq6}. Therefore, we only consider the contribution of gluons from the resolved photon for heavy quarkonium photoproduction in this work. According to the NRQCD factorization formalism, the total cross section for the $g_{\gamma}+p\rightarrow V+X\otimes p$ processes is calculated, which can be expressed as follows
\be
\sigma (g_{\gamma}+p \rightarrow V+X\otimes p)=\int dzdp_T^2dx_3\frac{x_2g_h(x_2,Q^2)x_3g_{\gamma}(x_3,Q^2)}{z(1-\frac{z}{x_3})}\frac{d\sigma}{dt}(g+g\rightarrow V+g),
\ee
where the variable $x_2$ is the momentum fraction of hadron carried by the gluon and $x_3$ is the momentum fraction of photon carried by the gluon in the resolved photoproduction processes. The variable $x_2$ can be written as
\be
x_2=\frac{x_3p_T^2+M^2(x_3-z)}{W_{\gamma p}^2z(1-\frac{z}{x_3})}.
\ee

The total cross section for $g_{\gamma}+p$ process is similar to the $\gamma +p$ process, but it has a term about the photon structure function $g_\gamma(x_3,Q^2)$. In our work, we use the parametrization of photonic parton distributions in Ref. \cite{photon structure}. In the $g_\gamma+p\rightarrow V+X\otimes p$ process the partonic differential cross section $d \sigma /dt$ is similar to the $\gamma+p \rightarrow V+X\otimes p$ process, which includes the contributions of color-singlet and color-octet \cite{gammah, octet:dsdt}. Based on the kinematics for the resolved photoproduction process ($g_\gamma +p\rightarrow V+X\otimes p$) the Mandelstam variables can be written as
\begin{eqnarray}
s &=& \frac{x_3p_T^2+M^2(x_3-z)}{z(1-\frac{z}{x_3})}, \notag \\
t &=& -\frac{x_3p_T^2+M^2(x_3-z)}{z}, \notag \\
u &=& M^2-\frac{zs}{x_3}.
\end{eqnarray}

\section{Numerical Results} \label{sec:result}
To explore the contribution of the resolved photoproduction processes, we firstly calculate the total cross section distribution of the inclusive $J/\Psi$ production and compare our results with the H1 data \cite{H1colla}. Subsequently we present the predictions of rapidity and transverse momentum distributions of the inclusive diffractive $J/\Psi$, $\Psi(2S)$ and $\Upsilon$ photoproduction in $pp$, $pPb$ and $ PbPb$ collisions. In this work, we only consider the NRQCD at leading order (LO). As shown in Ref. \cite{Artoisenet:2009xh} the results of next-to-leading order (NLO) have large corrections for the cross section of quarkonium, and the NLO corrections can reach $70\%$ in Ref. \cite{Kramer:1995nb}. Certainly, in the future the NLO corrections deserve detailed investigation for heavy quarkonium photoproduction at LHC energies.

In our calculations, we have set $Q^2=p_T^2$ and use the CTEQ6LO \cite{cteq6} parton distribution function (PDF) for the inclusive gluon distribution. The mass of charm quark is equal to 1.5 GeV and bottom quark's mass is equal to 4.5 GeV. Following Ref. \cite{H1colla}, the fraction of photon energy carried by the quarkonium is integrated over the range $0.3 \leq z\leq 0.9$, and the minimum value of $p_T$ is 1 GeV. In addition, the long-distance matrix elements of the color-singlet and color-octet for $J/\Psi$, $\Psi(2S)$ and $\Upsilon$ are given \cite{cai.prc}
\begin{eqnarray}
&&\Big \langle \mathcal{O}^{J/\Psi}\Big [\,^3S_1^{[1]}\Big] \Big \rangle=1.2 GeV^3, \notag \\
&&\Big \langle \mathcal{O}^{J/\Psi}\Big [\,^1S_0^{[8]}\Big] \Big \rangle=0.0180\pm 0.0087 GeV^3, \notag \\
&&\Big \langle \mathcal{O}^{\Psi(2S)}\Big [\,^3S_1^{[1]}\Big] \Big \rangle=0.76 GeV^3, \notag \\
&&\Big \langle \mathcal{O}^{\Psi(2S)}\Big [\,^1S_0^{[8]}\Big] \Big \rangle=0.0080\pm 0.0067 GeV^3, \notag \\
&&\Big \langle \mathcal{O}^{\Upsilon}\Big [\,^3S_1^{[1]}\Big] \Big \rangle=10.9 GeV^3, \notag \\
&&\Big \langle \mathcal{O}^{\Upsilon}\Big [\,^1S_0^{[8]}\Big] \Big \rangle=0.0121\pm0.0400 GeV^3.
\vspace{0.4cm}
\end{eqnarray}
\begin{figure}[t!]
\begin{center}
\vspace{0.0cm}\hspace{-1.9cm}
\includegraphics[scale=0.43]{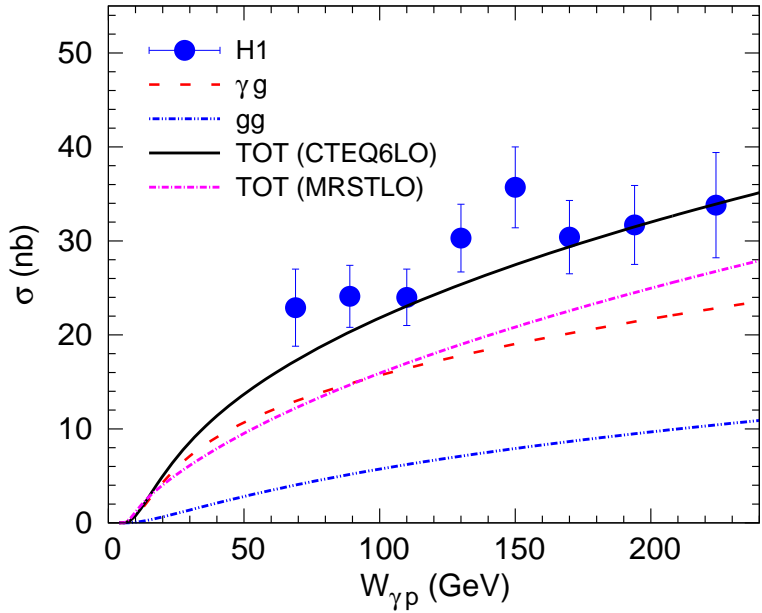}
\end{center}
\setlength{\abovecaptionskip}{0.1cm}
\setlength{\belowcaptionskip}{0.3cm}
\caption{The total cross sections ($\sigma$) as a function of the photon-proton center-of-mass energy ($W_{\gamma p}$) for the inclusive $J/\Psi$ production by direct ($\gamma g$) and resolved ($gg$) photoproduction processes in the $\gamma+p \rightarrow V+X\otimes p$ process. Data from H1 Collaboration \cite{H1colla}.}
\label{sigma_w}
\end{figure}

In Fig. \ref{sigma_w}, we show the energy dependence of the total cross sections which are calculated by considering the contribution of resolved photoproduction processes for the inclusive $J/\Psi$ production. The red dashed line is the $\gamma + g_p$ process, where the gluon emit from the proton. The blue dot-dot-dashed line is the $g_{\gamma} + g_p$ process, where the gluon emit from the resolved photon and the proton, respectively. The black solid line is the total contribution which includes photon-gluon and gluon-gluon processes. As we can see, our results are well consistent with the experimental data when the gluon-gluon process is included. Comparing the results with and without the contribution of gluon-gluon interaction, it is clear that the resolved photoproduction processes has a significant effect in heavy quarkonium inclusive photoproduction for the $\gamma+p \rightarrow V+X\otimes p$ process, and especially the contribution of resolved photoproduction processes in the region of larger $W_{\gamma p}$ can reach to about $28\%$.
\begin{figure}[t!]
\begin{center}
\vspace{-0.2cm}\hspace{-0.42cm}
\includegraphics[scale=0.232]{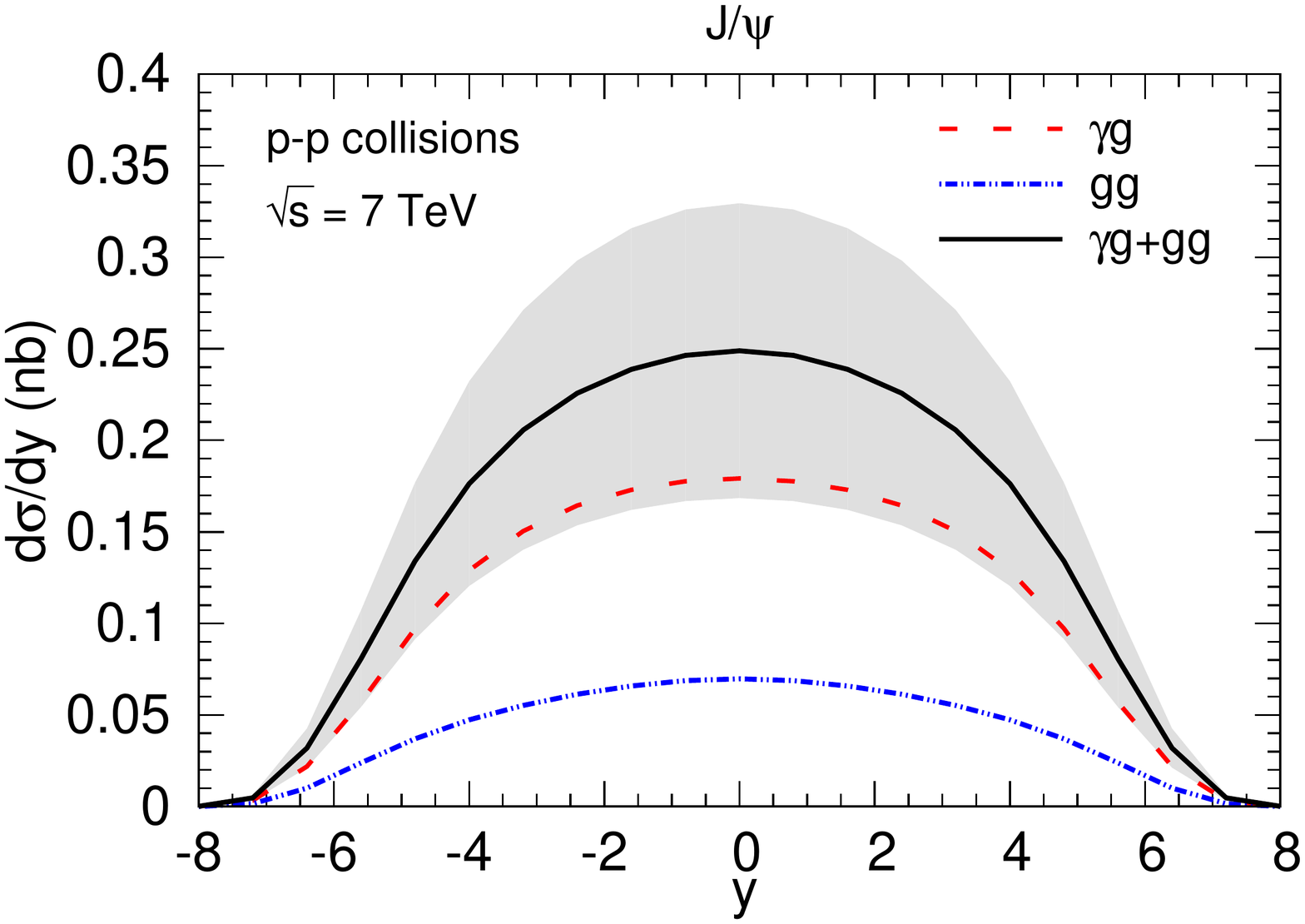}
\vspace{-0.2cm}\hspace{-1.30cm}
\includegraphics[scale=0.232]{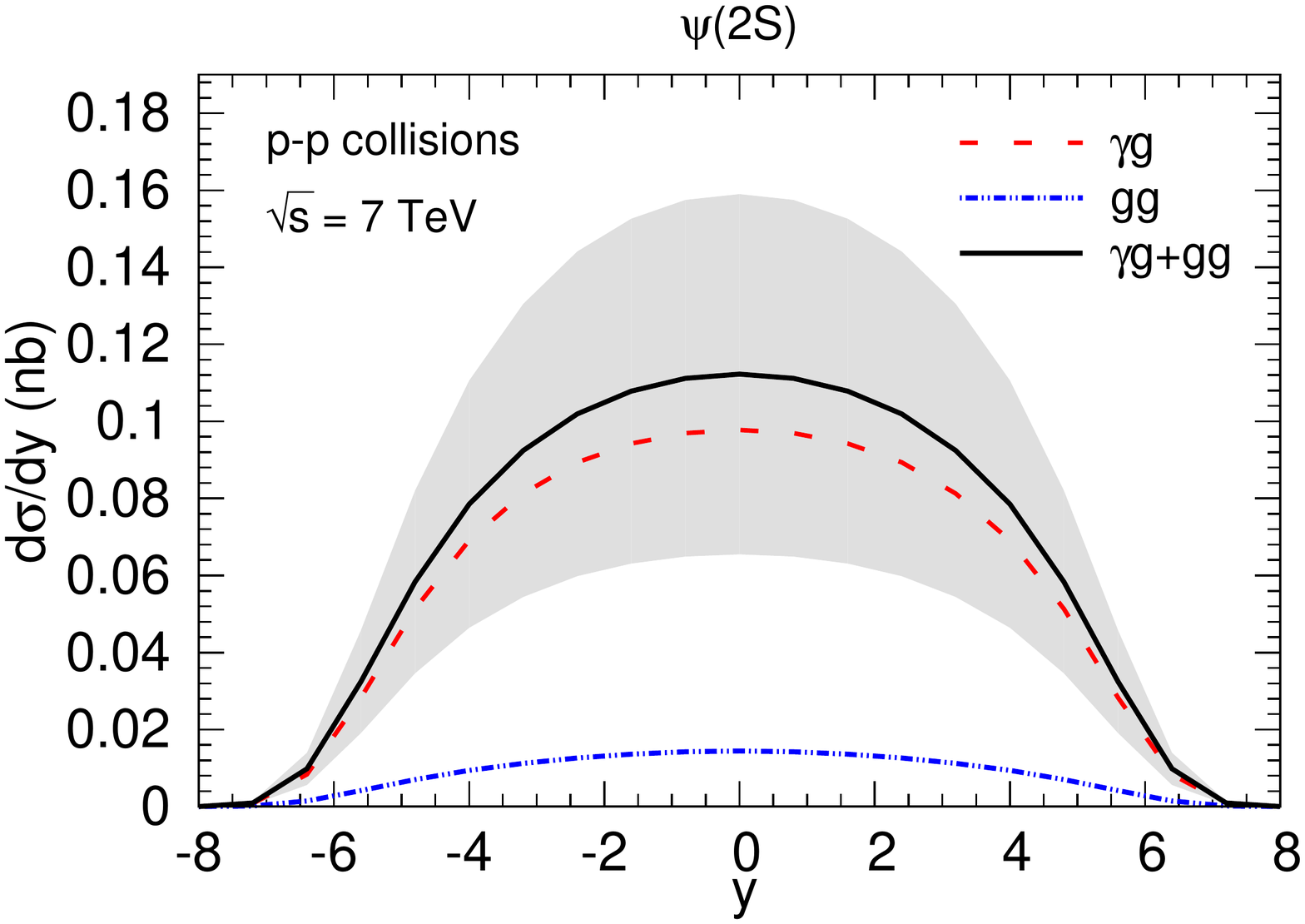}
\vspace{-0.2cm}\hspace{-1.42cm}
\includegraphics[scale=0.232]{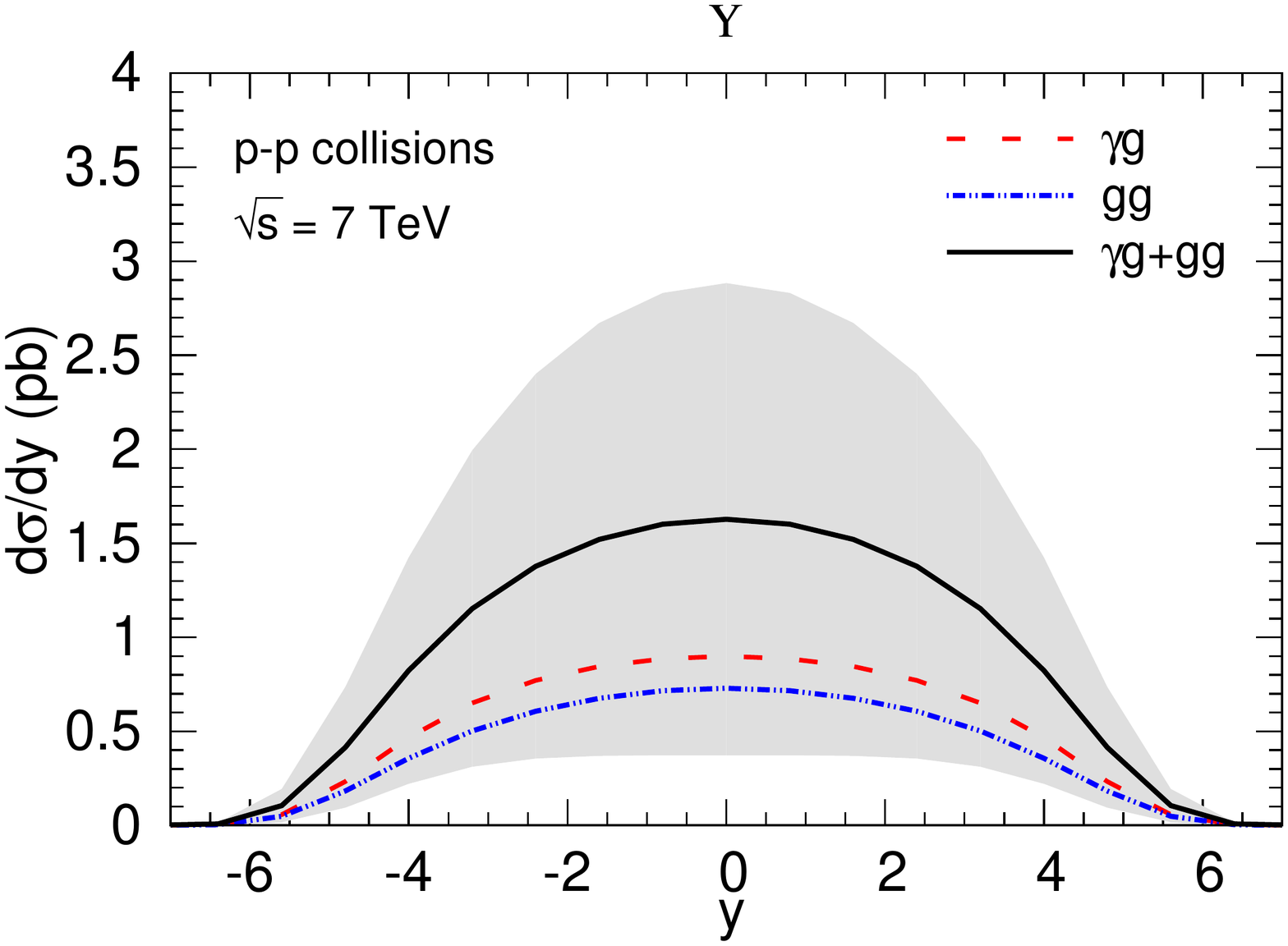}
\end{center}
\setlength{\abovecaptionskip}{0.1cm}
\setlength{\belowcaptionskip}{0.5cm}
\caption{Differential cross sections as a function of rapidity for the inclusive diffractive heavy quarkonium production by direct ($\gamma g$) and resolved ($gg$) photoproduction processes based on the resolved pomeron model in $pp$ collisions. The gray bands show the uncertainties for the total contribution ($\gamma g+gg$) associated with the theoretical uncertainty of the LDME and the scale $Q^2$ from $p_T^2/2$ to $2p_T^2$.}
\label{pp}
\end{figure}

\begin{table*}[t!]
\centering
\begin{tabular}{|c|c|c|c|c|c|c|}
\hline
~~~ - ~~~&$pp$ ($\sqrt{s}= 7$~TeV) & $pPb$ ($\sqrt{s}= 5$~TeV) & $PbPb$ ($\sqrt{s}= 5.5$~TeV)   \\ \hline
~~~$J/\Psi   $  ~~~&~~~ 1.59~-~3.11 nb  ~~~&~~~ 1.73~-~3.36 $\mu b$  ~~~&~~~ 0.24~-~0.45 mb   ~~~  \\ \hline
~~~$\Psi(2S) $  ~~~&~~~ 0.61~-~1.47 nb  ~~~&~~~ 0.62~-~1.49 $\mu b$  ~~~&~~~ 0.099~-~0.22 mb   ~~~  \\ \hline
~~~$\Upsilon $  ~~~&~~~ 3.08~-~21.93 pb ~~~&~~~ 2.20~-~14.77  nb     ~~~&~~~ 0.48~-~1.91 $\mu b$ ~~~ \\ \hline
\end{tabular}
\vspace{0.5cm}
\setlength{\abovecaptionskip}{0.4cm}
\setlength{\belowcaptionskip}{0.5cm}
\caption{The lower and upper bounds of the total cross sections associated with the LDME determined by the theoretical uncertainty of $^1S_0^{[8]}$ matrix element and the scale $Q^2$ from $p_T^2/2$ to $2p_T^2$ for the total contribution ($\gamma g+gg$).}
\label{tot_uncertainty}
\end{table*}

The $J/\Psi$ production includes the feed-down contributions from the higher charmonium states, which has not been computed in this work. The feed-down contribution from $\Psi(2S)$ is about $15\%$ for the cross section of $J/\Psi$ production \cite{H1colla,Chekanov:2002at}, and the contribution from $\chi_c$ is about $20\%$ \cite{LHCb:2012af}. Nevertheless these contributions are only important when $z$ is in the low region. Since $z$ in our calculations is in the range $0.3 \leq z\leq 0.9$, we consider only the direct contributions. In Fig. \ref{sigma_w} we also present the result obtained using the MRSTLO \cite{Pumplin:2002vw} parton distributions for the proton. By comparison with the result obtained using the CTEQ6LO, the result of MRSTLO is almost $20\%$ smaller at larger energies. In Fig. \ref{sigma_w} we only take into account the central values of the LDME and the scale $Q^2$, but the leading order NRQCD calculations suffer from large uncertainties.  As a consequence, in what follows we will present the uncertainty bands of the total contribution ($\gamma g+gg$) for the heavy quarkonium inclusive diffractive photoproduction, which associate with the LDME determined by the theoretical uncertainty of $^1S_0^{[8]}$ matrix element and the scale $Q^2$ from $p_T^2/2$ to $2p_T^2$. Meanwhile, we also present the lower and upper bounds of the total cross sections for the total contribution ($\gamma g+gg$) in Table \ref{tot_uncertainty}. As expected, the large uncertainties are presented.

In the inclusive $J/\Psi$ production, the resolved photoproduction processes play an important role. Analogously, we extend the resolved photoproduction processes to the resolved pomeron model to predict the rapidity distributions of the inclusive diffractive heavy quarkonium $J/\Psi$, $\Psi(2S)$ and $\Upsilon$ photoproduction. In Fig. \ref{pp}, we show the rapidity distributions of inclusive diffractive $J/\Psi$, $\Psi (2S)$ and $\Upsilon$ photoproduction in $pp$ collisions at $\sqrt{s} = 7$~TeV. The red dashed lines are the interaction between photon and gluon radiated from the pomeron. The blue dot-dot-dashed lines represent the interaction of two gluons radiated from the resolved photon and the pomeron, respectively. The black solid lines are the total contribution which includes $\gamma+ g_{_{\mathbb{P}}}$ and $g_{\gamma} + g_{_{\mathbb{P}}}$ interactions. As we know, according to the NRQCD factorization formalism the differential cross sections decrease with the heavy quarkonium's mass. So the differential cross sections of $\Upsilon$ photoproduction is smaller than the $J/\Psi$ and $\Psi(2S)$. In addition the results show that the rapidity distributions are symmetric about the central rapidities ($y = 0 $) in that the both event hadrons are sources of photons. The total cross sections of direct and resolved photoproduction processes are presented in Table \ref{central_value}, and we note that for $pp$ collisions the contribution of resolved photoproduction processes can reach to $28\%$, $13\%$ and $44\%$ for the rapidity distributions of heavy quarkonium $J/\Psi$, $\Psi(2S)$ and $\Upsilon$ inclusive diffractive photoproduction, respectively.
\begin{figure}[t!]
\begin{center}
\vspace{-0.2cm}\hspace{-0.6cm}
\includegraphics[scale=0.232]{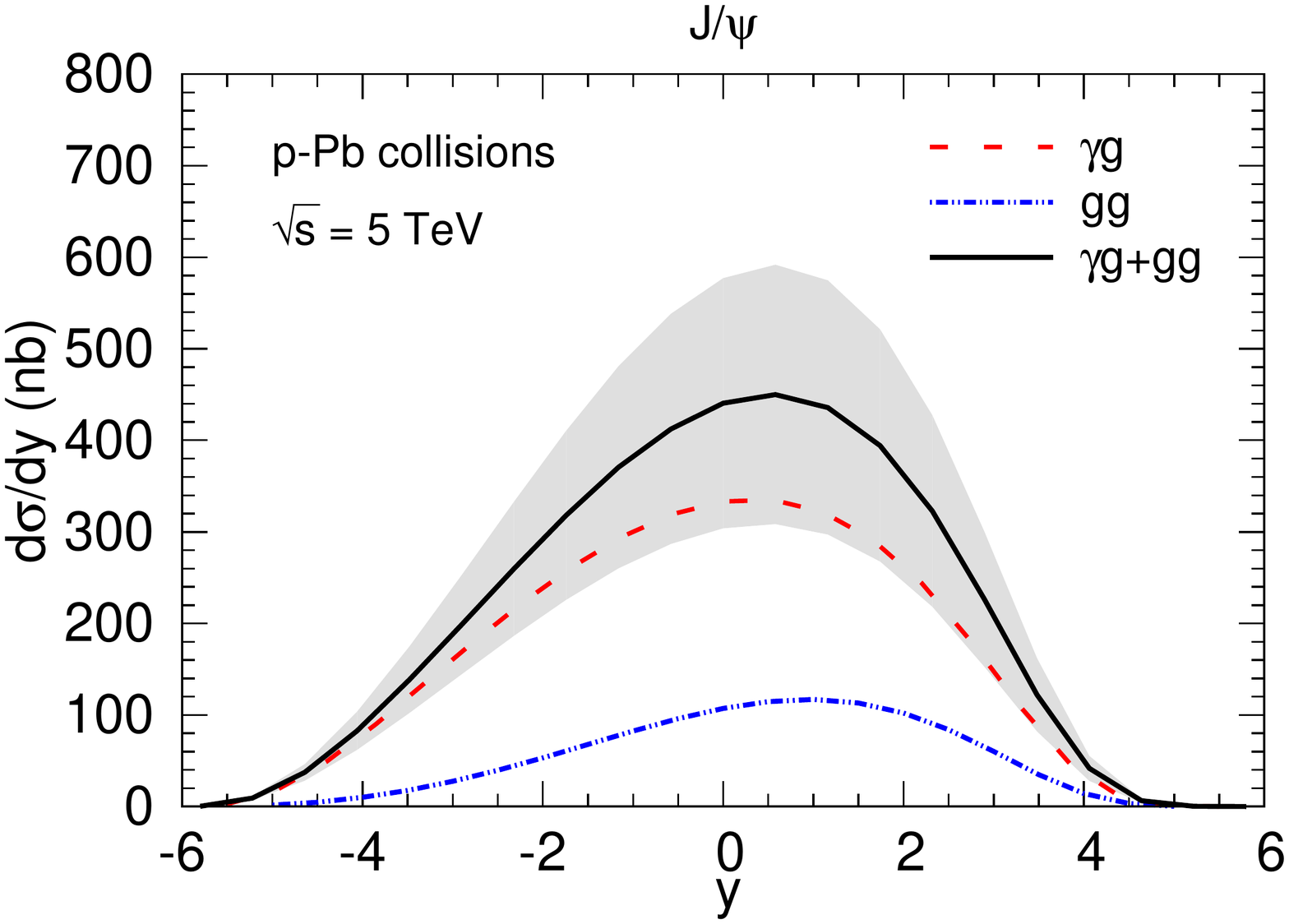}
\vspace{-0.4cm}\hspace{-1.30cm}
\includegraphics[scale=0.232]{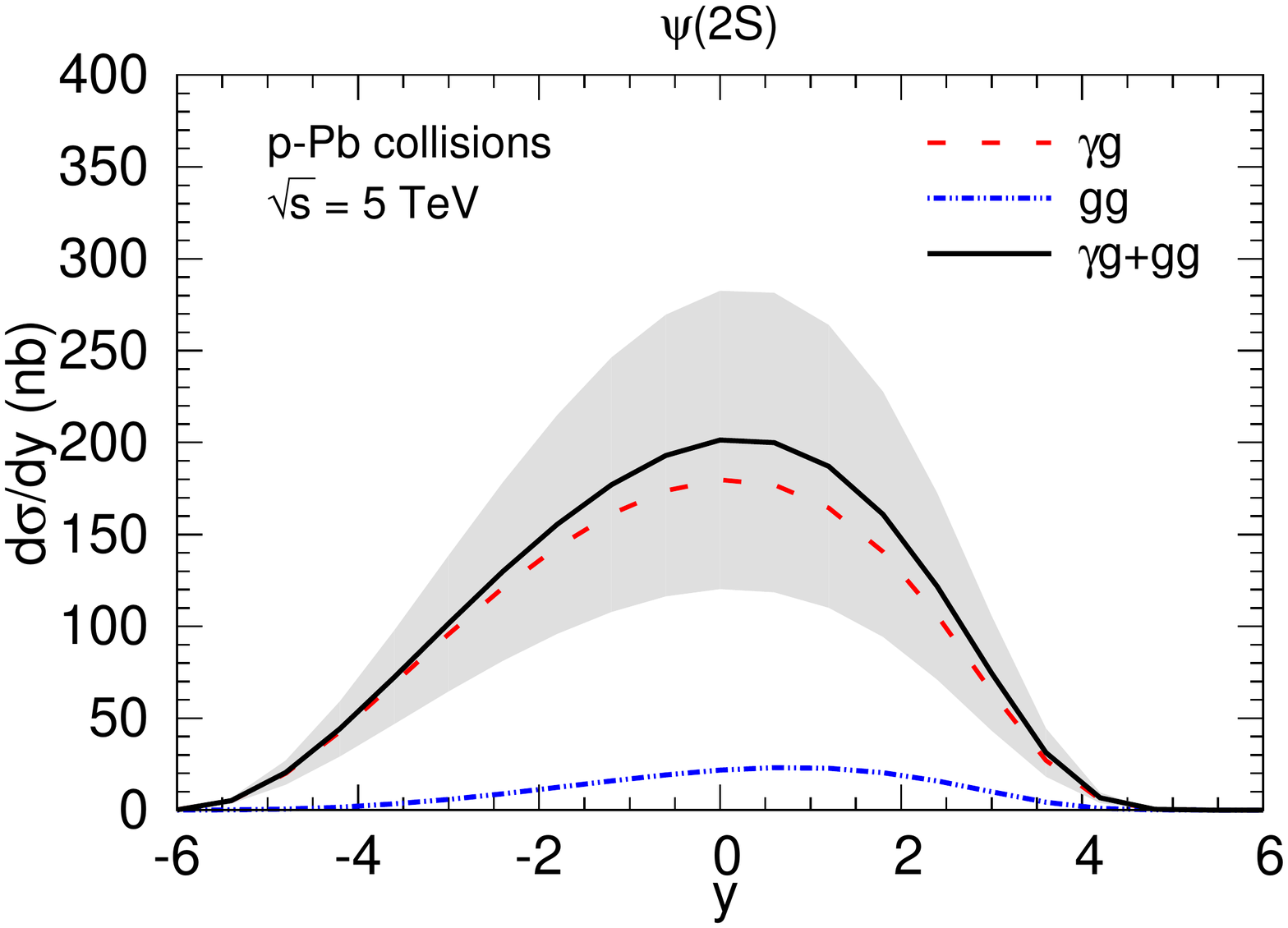}
\vspace{-0.4cm}\hspace{-1.54cm}
\includegraphics[scale=0.232]{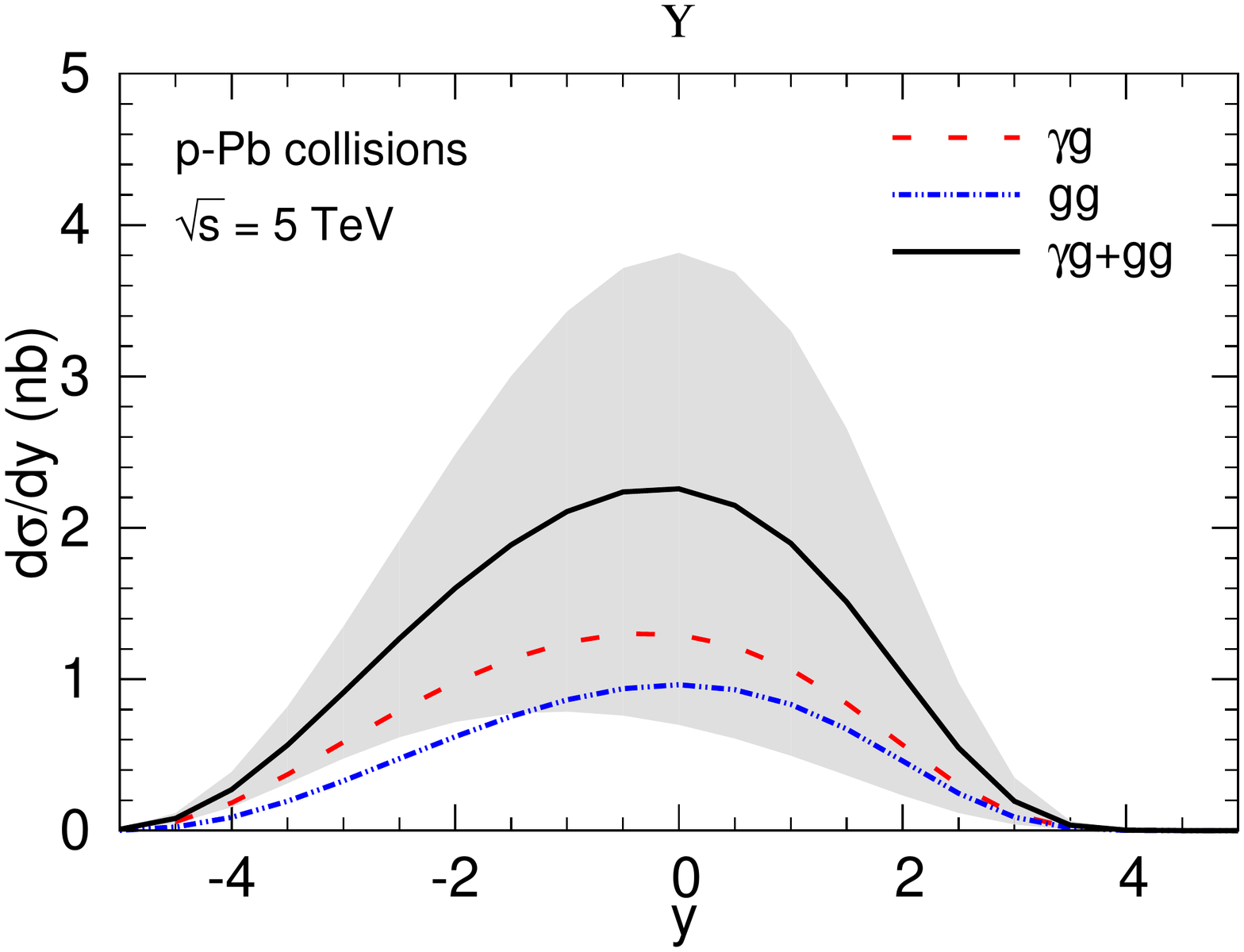}
\end{center}
\setlength{\abovecaptionskip}{0.1cm}
\setlength{\belowcaptionskip}{0.5cm}
\caption{Differential cross sections as a function of rapidity for the inclusive diffractive heavy quarkonium production by direct ($\gamma g$) and resolved ($gg$) photoproduction processes based on the resolved pomeron model in $pPb$ collisions. The gray bands show the uncertainties for the total contribution ($\gamma g+gg$) associated with the theoretical uncertainty of the LDME and the scale $Q^2$ from $p_T^2/2$ to $2p_T^2$.}
\label{ppb}
\end{figure}

\begin{table*}[t!]
\centering
\begin{tabular}{|c|c|c|c|c|c|c|}
\hline
-&\multicolumn{2}{|c|}{$pp$ ($\sqrt{s}= 7$~TeV)} & \multicolumn{2}{|c|}{$pPb$ ($\sqrt{s}= 5$~TeV)} & \multicolumn{2}{|c|}{$PbPb$ ($\sqrt{s}= 5.5$~TeV)}            \\ \hline
-             &  $\gamma g$   & $gg $    &  $ \gamma g$  & $gg$         & $\gamma g$               & $gg$                     \\ \hline
$~~~J/\Psi   $  ~~~&~~ 1.70 nb ~~&~~ 0.65 nb  ~~&~~ 1.86 $\mu b$ ~~&~~ 0.69 $\mu b$ ~~&~~ 0.27 mb       ~~&~~ 0.072 mb     ~~    \\ \hline
$~~~\Psi(2S) $  ~~~&~~ 0.91 nb ~~&~~ 0.13 nb  ~~&~~ 0.93 $\mu b$ ~~&~~ 0.13 $\mu b$ ~~&~~ 0.15 mb       ~~&~~ 0.014 mb     ~~     \\ \hline
$~~~\Upsilon $  ~~~&~~ 6.98 pb ~~&~~ 5.53 pb  ~~&~~ 4.76 nb      ~~&~~ 3.73 nb      ~~&~~ 0.72 $\mu b$  ~~&~~ 0.48 $\mu b$ ~~      \\ \hline
\end{tabular}
\vspace{0.5cm}
\setlength{\abovecaptionskip}{0.4cm}
\setlength{\belowcaptionskip}{0.5cm}
\caption{Total cross sections of inclusive diffractive heavy quarkonium $J/\Psi$, $\Psi(2S)$ and $\Upsilon$ direct ($\gamma g$) and resolved ($gg$) photoproduction in $pp$, $pPb$ and $PbPb$ collisions at the LHC energies.}
\label{central_value}
\end{table*}

As discussed before we only consider the $pp$ collisions, while the $pPb$ and $PbPb$ collisions at the LHC energies are still essential subjects \cite{Xie,Dillenseger}. Therefore, in what follows we will give predictions of the $pPb$ and $PbPb$ collisions. The equivalent photon flux of nucleus becomes \cite{nucleus photon spec}
\begin{eqnarray}
\frac{d N_{\gamma /A}(\omega )}{d\omega }= \frac{2\,Z^2\alpha_{em}}{\pi\,\omega}\, \left[\xi\,K_0\,(\xi)\, K_1\,(\xi) - \frac{\xi^2}{2}\,(K_1^2\,(\xi)-K_0^2\,(\xi)) \right],\,
\end{eqnarray}
where $K_0$ and $K_1$ are the modified Bessel functions and $Z$ is atomic number, with $\xi=\omega(R_{p_1}+R_{p_2})/\gamma_L$. In nucleus-nucleus processes, the diffractive gluon distribution can be expressed as follows \cite{pomeron in Pb}
\begin{eqnarray}
g_{A}(x,Q^2)=R_g \,A^2 \int_x^1 \frac{dx_{\mathbb{P}}}{x_{\mathbb{P}}} \Big[f_{\mathbb{P}/p}(x_{\mathbb{P}})\cdot F_A^2(t)\Big]g_{\mathbb{P}}\Big(\frac{x}{x_{\mathbb{P}}},Q^2\Big),
\end{eqnarray}
where $R_g = 0.15$ denotes the suppression factor associated to the nuclear shadowing and $F_{A}(t)\propto e^{R_{A}^2 t/6}$ is the nuclear form factor, with $R_{A}$ being the radius of nucleus. In Fig. \ref{ppb}, we present our predictions of the rapidity distributions for $J/\Psi$, $\Psi(2S)$ and $\Upsilon$ photoproduction in $pPb $ collisions at $\sqrt{s}= 5$~TeV. It shows that rapidity distributions are asymmetric about the central rapidities ($y = 0 $) because the factor $Z$ strengthens the contribution of nucleus in proton-nucleus collision. Moreover, from $\omega =(M/2) exp(\pm y)$ one can find that the contribution of nucleus is dominant in the positive rapidity region, while the contribution of the proton is dominant in the negative rapidity region. One can easily see, from the Table \ref{central_value}, the contribution of resolved photoproduction processes can reach respectively $27\%$, $12\%$ and $44\%$ for the rapidity distributions of heavy quarkonium $J/\Psi$, $\Psi(2S)$ and $\Upsilon$ inclusive diffractive photoproduction in $pPb $ collisions.

\begin{figure}[t!]
\begin{center}
\vspace{-0.2cm}\hspace{-0.7cm}
\includegraphics[scale=0.232]{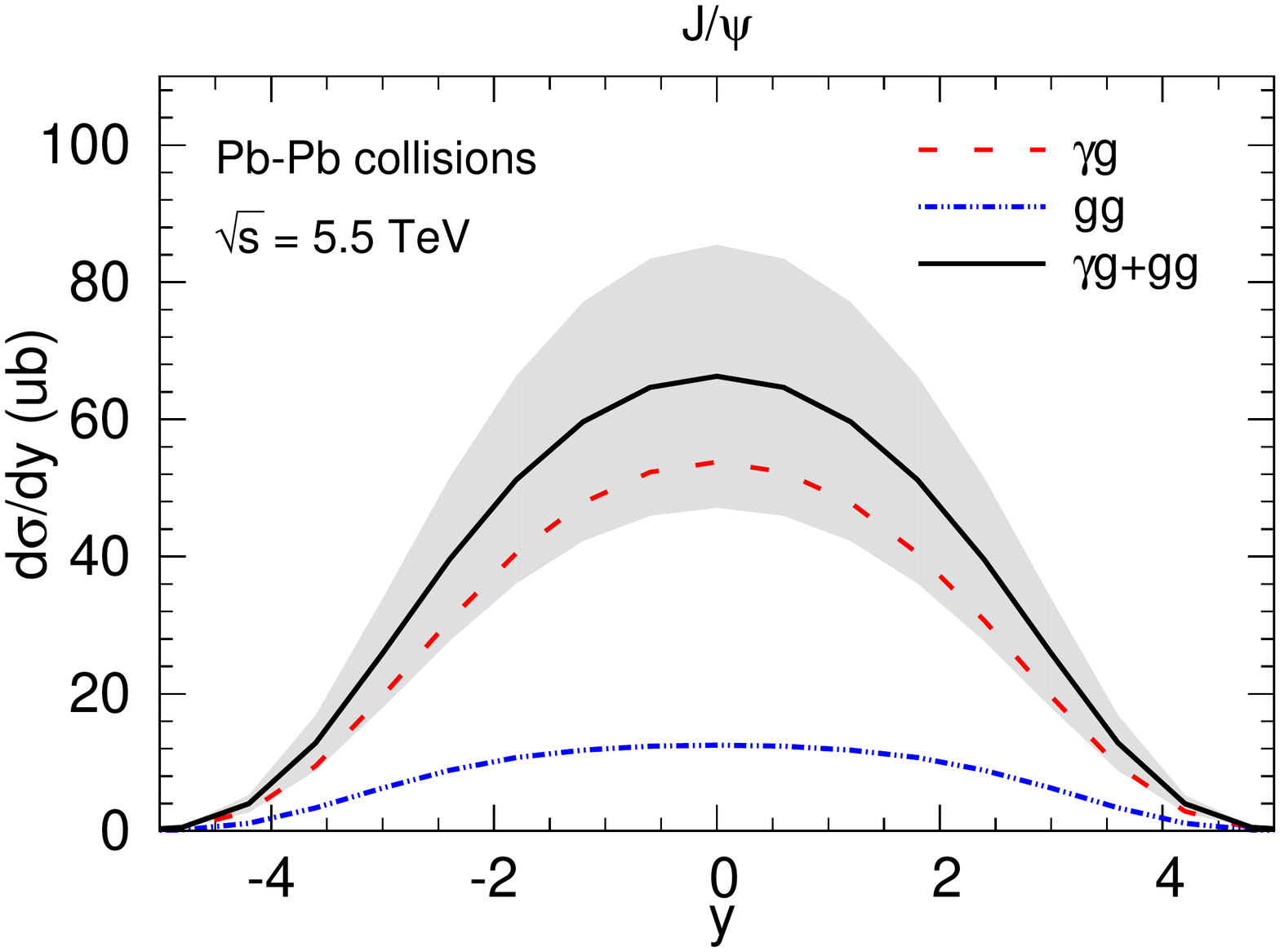}
\vspace{-0.4cm}\hspace{-1.39cm}
\includegraphics[scale=0.232]{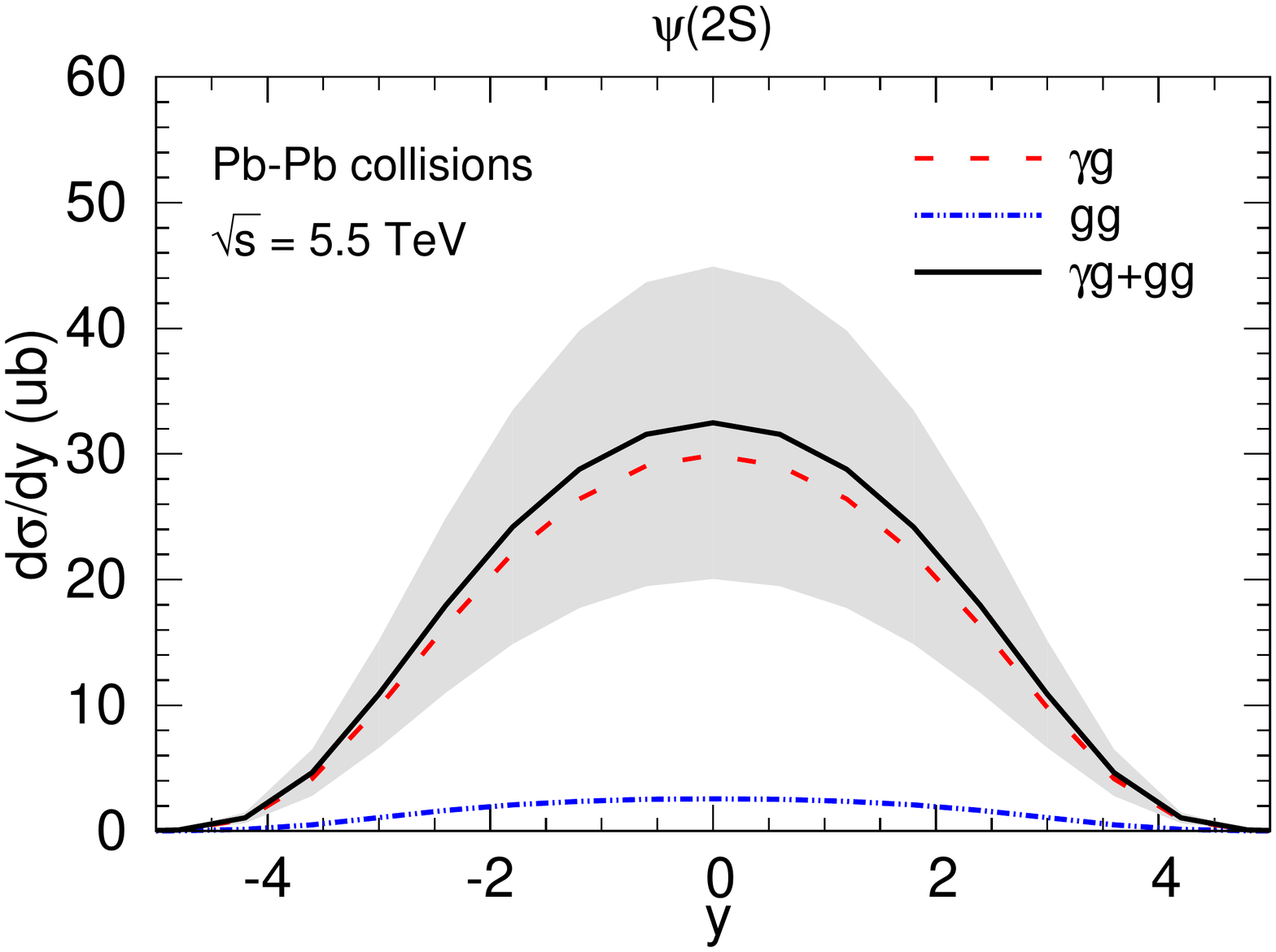}
\vspace{-0.4cm}\hspace{-1.43cm}
\includegraphics[scale=0.232]{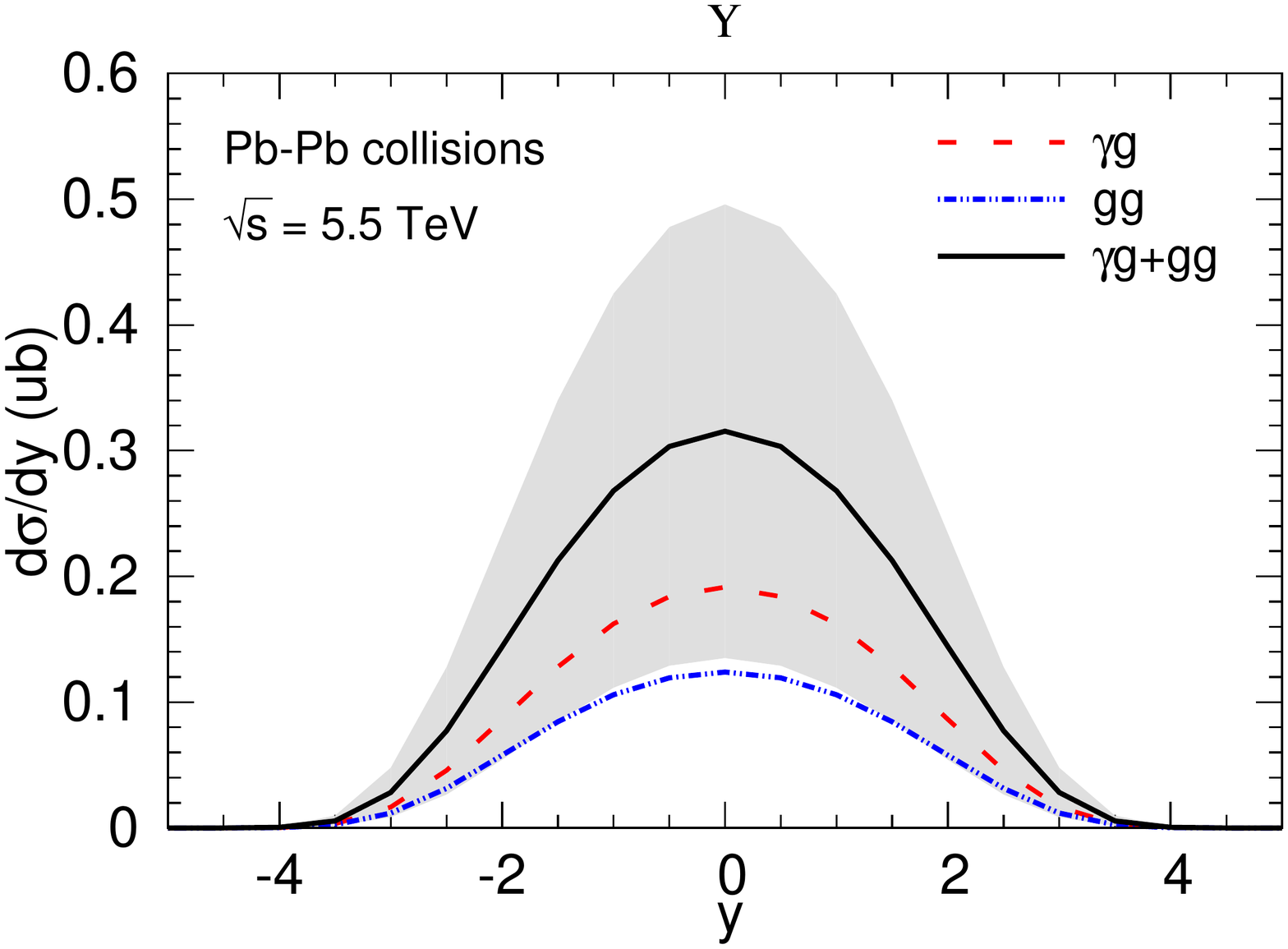}
\end{center}
\setlength{\abovecaptionskip}{0.1cm}
\setlength{\belowcaptionskip}{0.5cm}
\caption{Differential cross sections as a function of rapidity for the inclusive diffractive heavy quarkonium production by direct ($\gamma g$) and resolved ($gg$) photoproduction processes based on the resolved pomeron model in $PbPb$ collisions. The gray bands show the uncertainties for the total contribution ($\gamma g+gg$) associated with the theoretical uncertainty of the LDME and the scale $Q^2$ from $p_T^2/2$ to $2p_T^2$.}
\label{pbpb}
\end{figure}

In Fig. \ref{pbpb}, we present the rapidity distributions of inclusive diffractive heavy quarkonium photoproduction in $PbPb$ collisions at $\sqrt{s}=5.5$~TeV, and its total cross sections are presented in Table \ref{central_value}. It can be seen that the contributions of resolved photoproduction processes can reach to $21\%$, $8\%$ and $40\%$ for the rapidity distributions of heavy quarkonium $J/\Psi$, $\Psi(2S)$ and $\Upsilon$ inclusive diffractive photoproduction, respectively. Similar to the case of $pp$ and $pPb$ collisions, the contributions of resolved photoproduction processes for $\Upsilon$ production in $PbPb$ collisions are larger than $J/\Psi$ and $\Psi (2S)$. It is understandable since there is a charge factor $Q_q^2$ in differential cross section ($d \sigma /dt$) for $\gamma g$ process, but there is no such factor in differential cross section ($d \sigma /dt$) for $gg$ process. For charm quark the fractional electric charge is $Q_c=2/3$, but for bottom quark the fractional electric charge is $Q_b=-1/3$. In other words the $\gamma g$ process of charmonium production is 4 times as large as the $\gamma g$ process of bottomonium production. Therefore, the contribution of the $gg$ process will be relatively large for $\Upsilon$ production.

\begin{figure}[t!]
\begin{center}
\vspace{-0.2cm}\hspace{-0.5cm}
\includegraphics[scale=0.24]{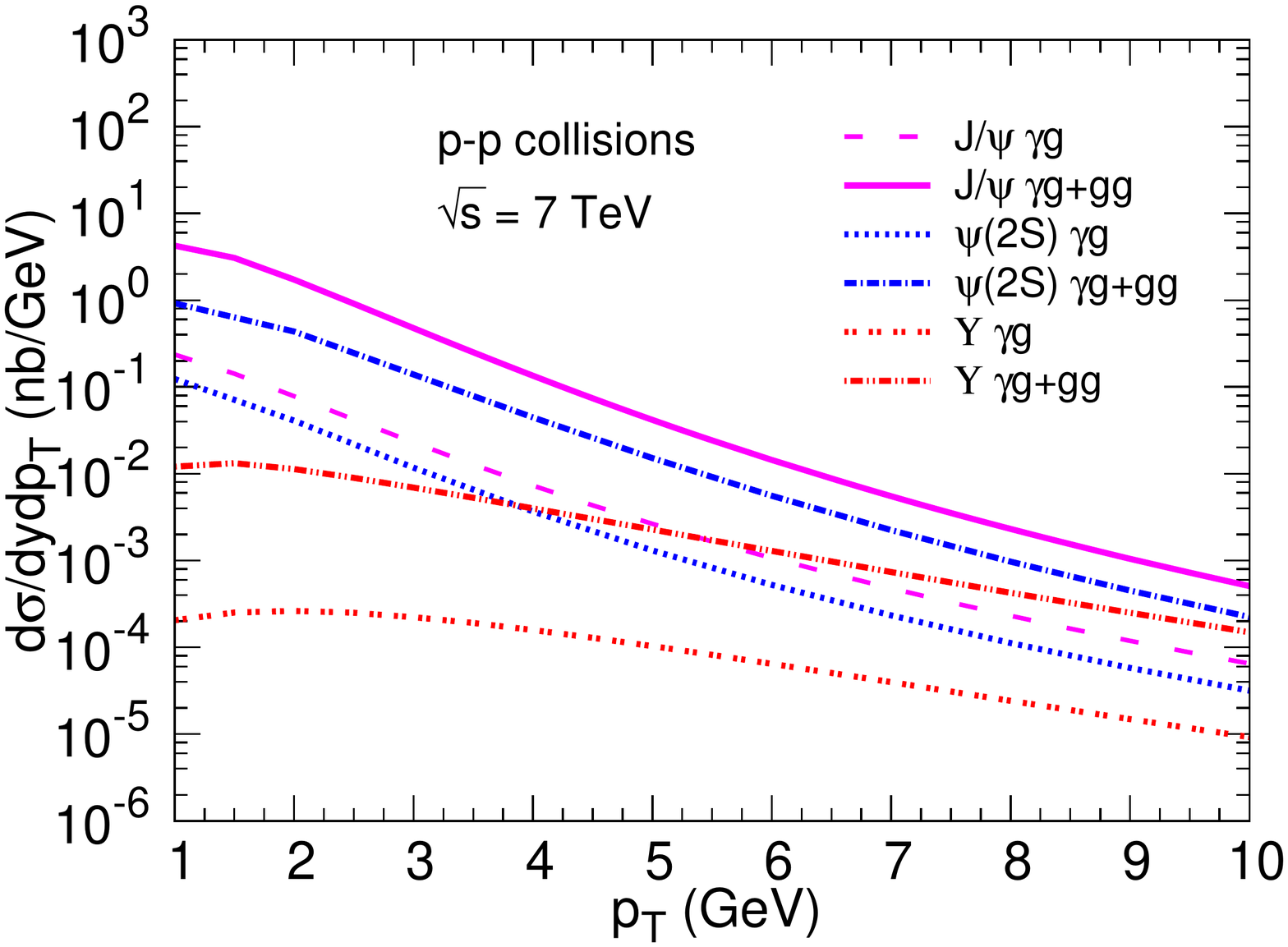}
\vspace{-0.2cm}\hspace{-1.65cm}
\includegraphics[scale=0.24]{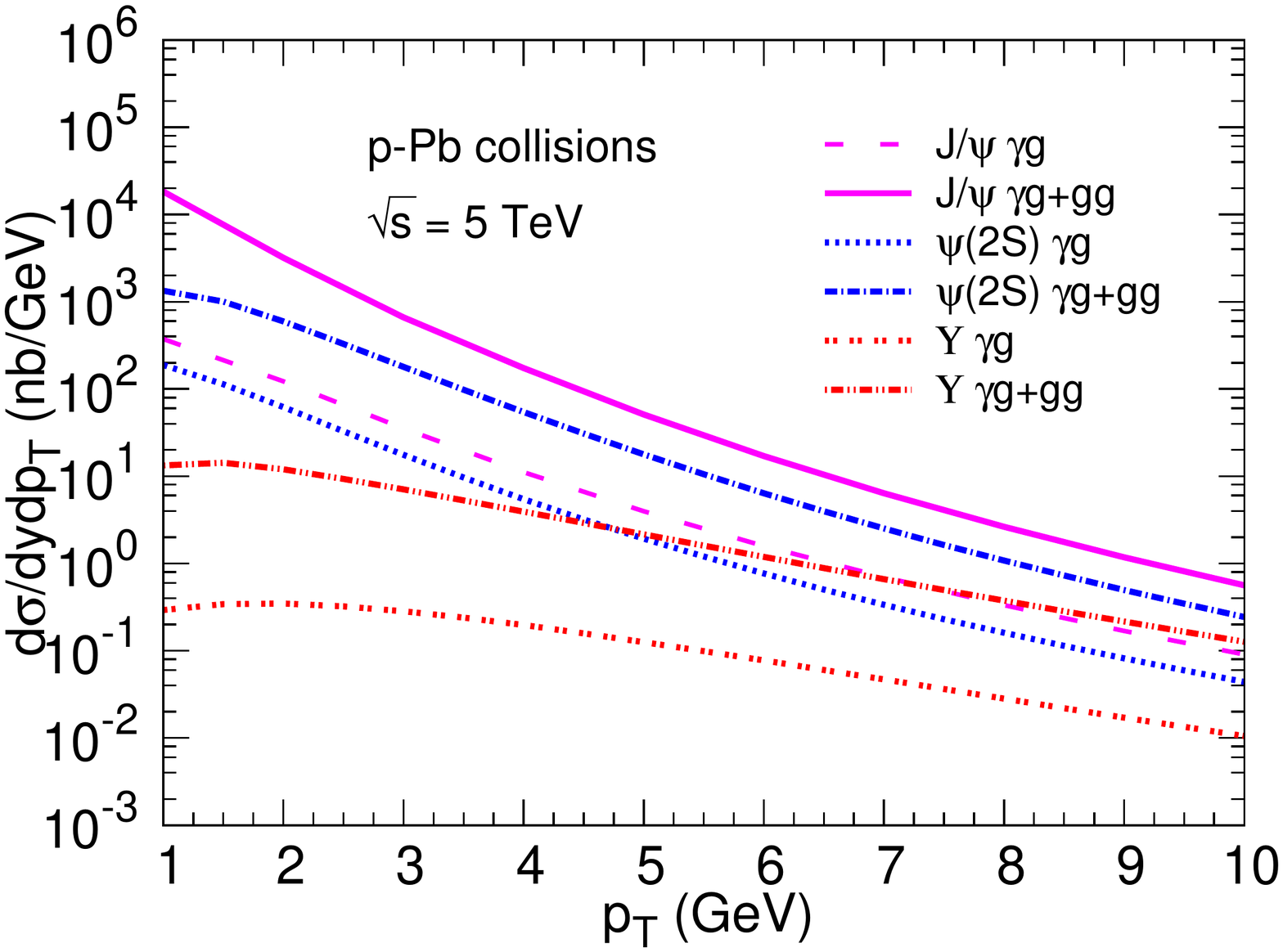}
\vspace{-0.2cm}\hspace{-1.65cm}
\includegraphics[scale=0.24]{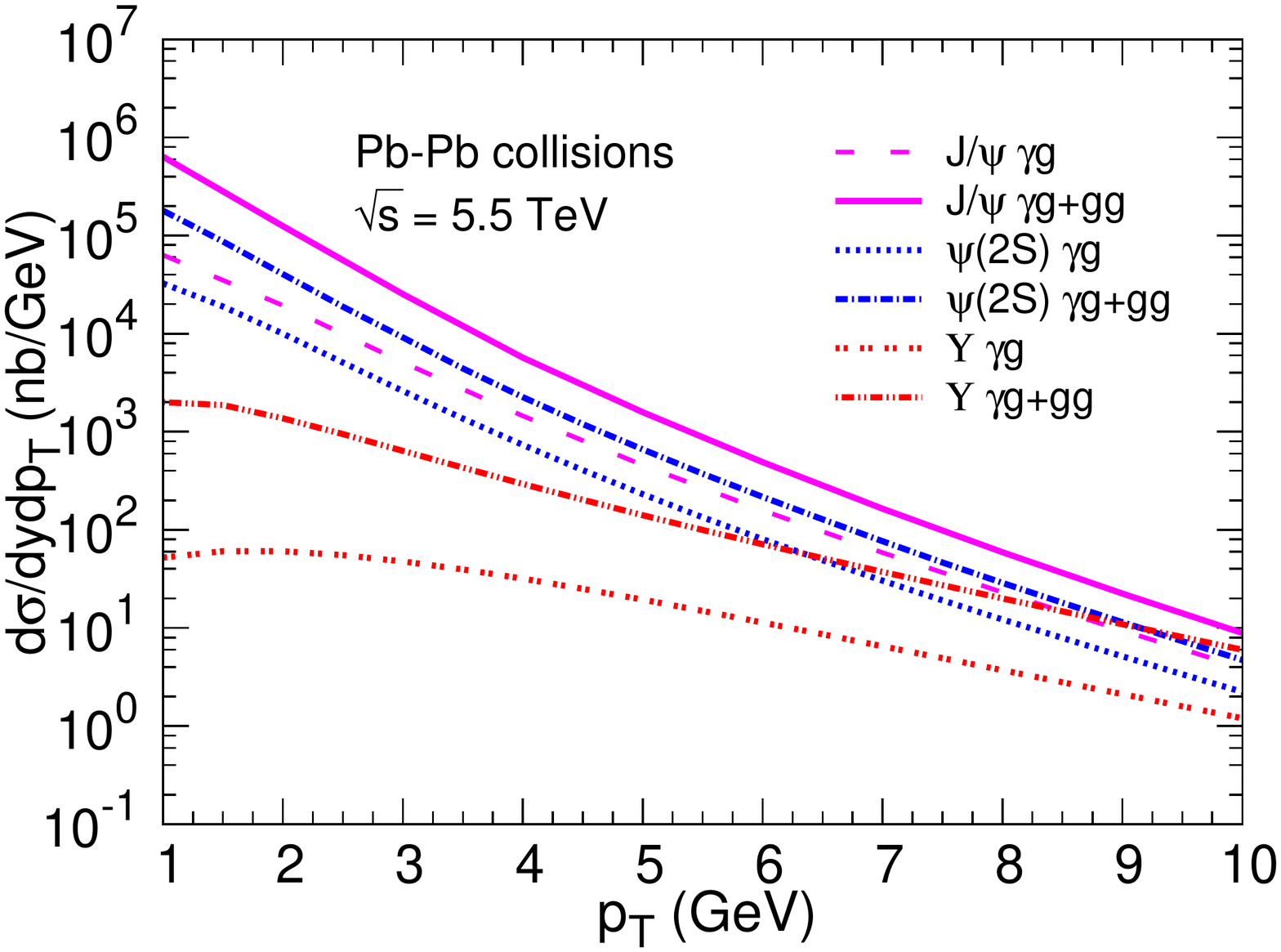}
\end{center}
\setlength{\abovecaptionskip}{0.1cm}
\setlength{\belowcaptionskip}{0.5cm}
\caption{Transverse momentum distributions for the inclusive diffractive heavy quarkonium production by direct ($\gamma g$) and resolved ($gg$) photoproduction processes based on the resolved pomeron model at central rapidities ($y=0$) in $pp$ (left panel), $pPb$ (central panel) and $PbPb$ (right panel) collisions.}
\label{pt}
\end{figure}

The transverse momentum distributions of the heavy quarkonium inclusive diffractive production by direct and resolved photoproduction processes at central rapidities (y=0) in $pp$, $pPb$ and $ PbPb$ collisions are presented in Fig. \ref{pt}. As expected due to the fact that the mass of $\Upsilon$ is larger than charmonium, the transverse momentum distributions of $\Upsilon$ are smaller than that of $J/\Psi$ and $\Psi (2S)$. Moreover, we note that the $gg$ process has notable contribution for transverse momentum distributions of heavy quarkonium $J/\Psi$, $\Psi (2S)$ and $\Upsilon$ inclusive diffractive photoproduction. Therefore, the resolved photoproduction processes are essential ingredient for the heavy quarkonium inclusive diffractive photoproduction.

\section{Summary} \label{sec:summary}
Using the NRQCD factorization formalism, we have studied the distributions of total cross sections of inclusive $J/\Psi$ production. The contribution of resolved photoproduction processes can reach to $28\%$ in the region of larger $W_{\gamma p}$, which indicates that we cannot ignore the contribution of resolved photoproduction processes in $\gamma p$ scattering at HERA. Then, we predict the rapidity and transverse momentum distributions of $J/\Psi$, $\Psi(2S)$ and $\Upsilon$ inclusive diffractive photoproduction in $pp$, $pPb$ and $PbPb$ collisions at LHC energies, which not only consider the resolved photoproduction precesses but also the resolved pomeron model. We note that the contribution of resolved photoproduction processes can separately reach $28\%$, $13\%$ and $44\%$ for the rapidity distributions of heavy quarkonium $J/\Psi$, $\Psi(2S)$ and $\Upsilon$ inclusive diffractive photoproduction in $pp$ collisions. For $pPb$ collisions the contribution of resolved photoproduction processes can reach to $27\%$, $12\%$ and $44\%$, respectively. For $PbPb$ collisions its contribution can reach to $21\%$, $8\%$ and $40\%$, respectively. These numerical results show that the resolved photoproduction processes are indispensable part for heavy quarkonium production, especially for $\Upsilon$ photoproduction.

\begin{acknowledgments}
This work is supported by the National Natural Science Foundation of China under Grants No. 11765005, and No. 11264005; Qian Kehe Platform Talents No. [2017]5736-027; the Department of Science and Technology of Guizhou Province under Grants No. [2018]1023, and No. [2019]5653; the Department of Education of Guizhou Province under Grant No. KY[2017]004; and the 2018 scientific research startup foundation
for the introduced talent of Guizhou University of Finance and Economics under grant No. 2018YJ60.
\end{acknowledgments}



\end{document}